\documentclass[12pt,preprint]{aastex}

\begin{document}

\def\green{f_{_{\rm G}}}
\def\Ngreen{N_{_{\rm G}}}
\def\zmin{z_{\rm min}}
\def\zmax{z_{\rm max}}
\def\xmin{x_{\rm min}}
\def\xmax{x_{\rm max}}

\slugcomment{submitted to \apj}

\shorttitle{Stochastic Particle Acceleration}
\shortauthors{Becker, \Le, \& Dermer}

\title{TIME-DEPENDENT STOCHASTIC PARTICLE ACCELERATION IN
ASTROPHYSICAL PLASMAS: \break EXACT SOLUTIONS INCLUDING MOMENTUM-DEPENDENT
ESCAPE}

\author{Peter A. Becker\altaffilmark{1}$^,$\altaffilmark{2}}

\affil{Center for Earth Observing and Space Research, \break
George Mason University \break
Fairfax, VA 22030-4444, USA}

\author{Truong Le\altaffilmark{3} \& Charles D. Dermer\altaffilmark{4}}
\affil{E. O. Hulburt Center for Space Research \break
Naval Research Laboratory, \break
Washington, DC 20375, USA}

\vfil

\altaffiltext{1}{pbecker@gmu.edu}
\altaffiltext{2}{also Department of Physics and Astronomy,
George Mason University, Fairfax, VA 22030-4444, USA}
\altaffiltext{3}{truong.le@nrl.navy.mil}
\altaffiltext{4}{dermer@gamma.nrl.navy.mil}

\begin{abstract}
Stochastic acceleration of charged particles due to interactions with
magnetohydrodynamic (MHD) plasma waves is the dominant process leading
to the formation of the high-energy electron and ion distributions in a
variety of astrophysical systems. Collisions with the waves influence
both the energization and the spatial transport of the particles, and
therefore it is important to treat these two aspects of the problem in a
self-consistent manner. We solve the representative Fokker-Planck
equation to obtain a new, closed-form solution for the time-dependent
Green's function describing the acceleration and escape of relativistic
ions interacting with Alfv\'en or fast-mode waves characterized by
momentum diffusion coefficient $D(p)\propto p^q$ and mean particle
escape timescale $t_{\rm esc}(p) \propto p^{q-2}$, where $p$ is the
particle momentum and $q$ is the power-law index of the MHD wave
spectrum. In particular, we obtain solutions for the momentum
distribution of the ions in the plasma and also for the momentum
distribution of the escaping particles, which may form an energetic
outflow. The general features of the solutions are illustrated via
examples based on either a Kolmogorov or Kraichnan wave spectrum. The
new expressions complement the results obtained by Park and Petrosian,
who presented exact solutions for the hard-sphere scattering case
($q=2$) in addition to other scenarios in which the escape timescale has
a power-law dependence on the momentum. Our results have direct
relevance for models of high-energy radiation and cosmic-ray production
in astrophysical environments such as $\gamma$-ray bursts, active galaxies,
and magnetized coronae around black holes. In particular, we outline an
application of the new results to black-hole sources that produce
outflows of relativistic hadrons, with associated predictions that can
be tested using GLAST.

\end{abstract}



\section{INTRODUCTION}

Observations of high-energy radiation from a variety of astrophysical
sources imply the presence of significant populations of nonthermal
(often relativistic) particles. Much of the observational data is
characterized by strong variability on very short timescales. Nonthermal
distributions are naturally produced via the Fermi process, in which
particles interact with scattering centers moving systematically and/or
randomly. The first-order Fermi mechanism treats particle acceleration
in converging flows, such as shocks, and is thought to be important at
the Earth's bow shock, in certain classes of solar flares, for cosmic-ray
acceleration by supernova remnants, and in sources with relativistic
outflows, including blazars and $\gamma$-ray bursts \citep[for reviews,
see][]{be87,kir94}. The second-order process, as it was originally
conceived by Fermi, involved the stochastic acceleration of particles
scattering with randomly moving magnetized clouds \citep{fer49}. Later
refinements of this idea replaced the magnetized clouds with
magnetohydrodynamic (MHD) waves \citep[e.g.,][]{mel80}. The
second-order, stochastic Fermi process now finds application in a wide
range of astrophysical settings, including solar flares
\citep{mgr90,lpm04a}, clusters of galaxies \citep{sst87,pet01,bru04},
the Galactic center \citep{lpm04b,ad04}, and $\gamma$-ray bursts
\citep{wax95,dh01}.

The standard approach to modeling the acceleration of nonthermal
particles via interactions with MHD waves involves obtaining the
solution to a steady-state transport equation that incorporates
treatments of systematic and stochastic Fermi acceleration, radiative
losses, and particle escape \citep[e.g.,][]{sch84,ss89,lmp06}. However,
the prevalence of variability on short timescales in many sources calls
into question the validity of the steady-state interpretation.
\citet{pp95} provided a comprehensive review of the various
time-dependent solutions that have been derived in the past 30 years. In
most cases, it is assumed that the momentum diffusion coefficient,
$D(p)$, and the mean particle escape timescale, $t_{\rm esc}(p)$, have
power-law dependences on the particle momentum $p$. Although the set of
existing solutions covers a broad range of possible values for the
associated power-law indices, there are certain physically interesting
cases for which no analytical solution is currently available. For
example, \citet{dml96} have shown that for the stochastic acceleration
of relativistic particles due to resonant interactions with plasma waves
in black-hole magnetospheres, one obtains $D(p) \propto p^q$ and $t_{\rm
esc}(p) \propto p^{q-2}$, where $q$ is the index of the wavenumber
distribution (see \S~2). The analytical solution for the time-dependent
Green's function in this situation is of particular physical interest,
but it has not appeared in the previous literature. This has motivated
us to reexamine the associated Fokker-Planck transport equation for this
case, and to obtain a new family of closed-form solutions for the
secular Green's function. The resulting expression, describing the
evolution of an initially monoenergetic particle distribution,
complements the set of solutions discussed by \citet{pp95}. Our primary
goal in this paper is to present a detailed derivation of the exact
analytical solution and to demonstrate some of its key properties.
Detailed applications of our results to the modeling of particle
acceleration in black-hole accretion coronae and other astrophysical
environments will be presented in a separate paper.

The remainder of the paper is organized as follows. In \S~2 we review
the fundamental equations governing the acceleration of charged
particles interacting with plasma waves. The transport equation is
solved in \S~3 to obtain the time-dependent Green's function, and
illustrative results are presented in \S~4. The astrophysical implications
of our work are summarized in \S~5, and additional mathematical details are
provided in the Appendix.

\section{FUNDAMENTAL EQUATIONS}

Charged particles in turbulent astrophysical plasmas are expected to be
accelerated via interactions with whistler, fast-mode, and Alfv\'en
waves propagating in the magnetized gas. Here we consider a simplified
isotropic description of the wave energy distribution, denoted by
$W(k)$, where $W(k) dk$ represents the energy density due to waves with
wavenumber between $k$ and $k+dk$. The transport formalism we consider
assumes a power-law distribution for the wave-turbulence spectrum, which
implies definite relations between the momentum diffusion coefficient
and the momentum-dependent escape timescale
\citep[e.g.,][]{mel74,sch89a,sch89b,dml96}. MHD waves injected over a
narrow range of wavenumber cascade to larger wavenumbers, forming a
Kolmogorov or Kraichnan power spectrum over the inertial range with
$W(k) \propto k^{-q}$, where $q=5/3$ and $q=3/2$ for the Kolmogorov and
Kraichnan cases, respectively \citep[e.g.,][]{zm90}. The specific forms
we will adopt for the transport coefficients in \S~2.2 are based on the
physics of the resonant scattering processes governing the wave-particle
interactions \citep[see][]{dml96}. We assume that the nonthermal
particle distribution is isotropic, and focus on a detailed treatment of
the propagation of particles in momentum space due to wave-particle
interactions. The spatial aspects of the transport (i.e., the
confinement of the particles in the acceleration region) will be treated
in an approximate manner using a momentum-dependent escape term.

\subsection{Transport Equation}

The fundamental transport equation describing the propagation of
particles in the momentum space can be written in the flux-conservation
form \citep[e.g.,][]{bec92,sch89a,sch89b}
\begin{equation}
{\partial f \over \partial t}
= - {1 \over p^2} {\partial \over \partial p}\left\{
p^2 \left[
A(p) \, f - D(p) {\partial f \over \partial p}\right]
\right\}
- {f \over t_{\rm esc}(p)}
+ {S(p,t) \over 4 \pi p^2}
\ ,
\label{eq1}
\end{equation}
where $p$ is the particle momentum, $f(p,t)$ is the particle
distribution function, $D(p)$ denotes the momentum diffusion
coefficient, $t_{\rm esc}(p)$ is the mean escape time, $S(p,t)$
represents particle sources, and $A(p)$ describes any additional,
systematic acceleration or loss processes, such as shock acceleration or
synchrotron/inverse-Compton emission. The quantity in square brackets in
equation~(\ref{eq1}) describes the flux of particles through the
momentum space \citep{tnj71}, and the source term is defined so that
$S(p,t) \, dp \, dt$ gives the number of particles injected into the
plasma per unit volume between times $t$ and $t + dt$ with momenta
between $p$ and $p + dp$. The total particle number density $n(t)$ and
energy density $U(t)$ of the distribution $f(p,t)$ are computed using
\begin{equation}
n(t) = \int_0^\infty 4 \pi \, p^2 \, f(p,t) \, dp
\ , \ \ \ \ \ 
U(t) = \int_0^\infty 4 \pi \, \epsilon \, p^2 \, f(p,t)
\, dp \ ,
\label{eq2}
\end{equation}
where the particle kinetic energy $\epsilon$ is related to the Lorentz factor
$\gamma$ and the particle momentum $p$ by
\begin{equation}
\epsilon = (\gamma-1) \, m c^2 \ , \ \ \ \ \ 
\gamma=\left({p^2 \over m^2 c^2} + 1\right)^{1/2} \ ,
\label{eq3}
\end{equation}
and $m$ and $c$ denote the particle rest mass and the speed of light,
respectively.

Rather than solve equation~(\ref{eq1}) directly to determine $f(p,t)$
for a given source term $S(p,t)$, it is more instructive to first solve
for the Green's function, $\green(p_0,p,t_0,t)$, which satisfies the
equation
\begin{equation}
{\partial \green \over \partial t}
= - {1 \over p^2} {\partial \over \partial p}\left\{
p^2 \left[A(p) \, \green
- D(p) {\partial \green \over \partial p}
\right]\right\}
- {\green \over t_{\rm esc}(p)}
+ {\delta(p-p_0) \, \delta(t-t_0) \over 4 \pi p_0^2}
\ ,
\label{eq4}
\end{equation}
where $p_0$ is the initial momentum and $t_0$ is the initial time.
The source term in this equation corresponds to the injection of a single
particle per unit volume with momentum $p_0$ at time $t_0$. The particle
number and energy densities associated with the Green's function are
given by
\begin{equation}
n_{_{\rm G}}(t) = \int_0^\infty 4 \pi \, p^2 \,
\green(p_0,p,t_0,t) \, dp
\ , \ \ \ \ \ 
U_{_{\rm G}}(t) = \int_0^\infty 4 \pi \, \epsilon \, p^2 \,
\green(p_0,p,t_0,t) \, dp
\ .
\label{eq5}
\end{equation}
Once the Green's function solution has been determined, the {\it
particular solution} associated with an arbitrary source distribution
$S(p,t)$ can be computed using the integral convolution
\citep[e.g.,][]{bec03}
\begin{equation}
f(p,t) = \int_0^t \int_0^\infty
\green(p_0,p,t_0,t) \,
S(p_0,t_0) \, dp_0 \, dt_0
\ ,
\label{eq6}
\end{equation}
where we have assumed that the particle injection begins at time $t=0$
and no particles are present in the plasma for $t < 0$.

\subsection{Transport Coefficients}

In Appendix~A.1, we demonstrate that for
arbitrary particle energies, the mean rate of change of the
particle momentum due to stochastic acceleration is related to the
momentum diffusion coefficient $D(p)$ via
\begin{equation}
\Big\langle {dp \over dt} \Big\rangle \Big|_{\rm stoch} = {1 \over p^2}
{d \over dp}\left(p^2 D\right)
\ .
\label{eq8}
\end{equation}
The corresponding result for the mean rate of change of the kinetic energy
is \citep[see Appendix~A.2 and][]{mr89}
\begin{equation}
\Big\langle {d\epsilon \over dt} \Big\rangle \Big|_{\rm stoch}
= {1 \over p^2} {d \over dp} \left(p^2 v D \right) \ ,
\label{eq8b}
\end{equation}
where $v$ is the particle speed. If the MHD wave spectrum has the
power-law form $W \propto k^{-q}$ associated with Alfv\'en and fast-mode
waves, then the momentum dependences of the diffusion coefficient $D(p)$
and the mean escape time $t_{\rm esc}(p)$ describing the resonant
pitch-angle scattering of relativistic particles are given by
\citep[e.g.,][]{dml96,mr89}
\begin{equation}
D(p) = D_* \, m^2 c^2 \left({p \over m c}
\right)^q \ , \ \ \ \ \ \ \
t_{\rm esc}(p) = t_* \left({p \over m c}
\right)^{q-2} \ ,
\label{eq7}
\end{equation}
where $D_* \propto {\rm s}^{-1}$ and $t_* \propto {\rm s}$ are
constants. We shall focus on transport scenarios with $q \le 2$, so that
$t_{\rm esc}$ is either a decreasing or constant function of the
momentum $p$. In order to maintain the physical validity of the
escape-timescale approach used here, we must require that $t_{\rm esc}$
exceed the light-crossing time $L/c$ for a source with size $L$. This
implies a fundamental upper limit to the particle momentum when $q < 2$.

By combining equations~(\ref{eq8}) and (\ref{eq7}), we find that the
mean rate of change of the momentum for relativistic particles
accelerated stochastically by MHD waves is given by
\begin{equation}
\Big\langle {dp \over dt} \Big\rangle\Big|_{\rm stoch} = (q+2) \, D_*
\, m c \left({p \over m c}\right)^{q-1}
\ .
\label{eq9}
\end{equation}
For simplicity, we shall assume that the momentum dependence of the
additional, systematic loss/acceleration processes appearing in the transport
equation (\ref{eq1}), described by the coefficient $A(p)$, mimics that of the
stochastic acceleration (eq.~[\ref{eq9}]). We therefore write
\begin{equation}
A(p) = A_* \, m c \left({p \over m c}\right)^{q-1}
\ ,
\label{eq10}
\end{equation}
where the constant $A_* \propto {\rm s}^{-1}$ determines the positive
(negative) rate of systematic acceleration (losses). Note that this
formulation precludes the treatment of loss processes with a quadratic
energy dependence (e.g., inverse-Compton or synchrotron) since that
would imply $q=3$, which is outside the range considered here. However,
first-order Fermi acceleration at a shock or energy losses due to
Coulomb collisions can be treated by setting $q=2$ with $A_*$ either
positive or negative, respectively. This suggests that the results
obtained here are relevant primarily for the transport of energetic
ions. However, even in this application one needs to bear in mind that
synchrotron and inverse-Compton losses will become dominant at
sufficiently high energies \citep[e.g.,][]{sch84,ss89,lmp06}.

It is convenient to transform to the dimensionless momentum and time variables
$x$ and $y$, defined by
\begin{equation}
x \equiv {p \over m c} \ , \ \ \ \ \ \ \ 
y \equiv D_* \, t
\ ,
\label{eq11}
\end{equation}
in which case the transport equation~(\ref{eq4}) for the Green's function becomes
\begin{equation}
{\partial \green \over \partial y}
= {1 \over x^2} {\partial \over \partial x}\left(
x^{2+q} \, {\partial \green \over \partial x}\right)
- {a \over x^2} {\partial \over \partial x} \left(
x^{1+q} \, \green\right) - \theta \, x^{2-q} \, \green
+ {\delta(x-x_0) \, \delta(y-y_0) \over 4 \pi m^3 c^3 x_0^2}
\ ,
\label{eq12}
\end{equation}
where
\begin{equation}
x_0 \equiv {p_0 \over m c} \ , \ \ \ \ \ \ \ 
y_0 \equiv D_* \, t_0
\ ,
\label{eq13}
\end{equation}
and we have introduced the dimensionless constants
\begin{equation}
a \equiv {A_* \over D_*} \ , \ \ \ \ \ \ \ 
\theta \equiv {1 \over D_* \, t_*}
\ .
\label{eq14}
\end{equation}
Note that $x$ equals the particle Lorentz factor in applications involving
ultrarelativistic particles. The constant $a$ describes the relative
importance of systematic gains or losses compared with the stochastic
process.

\subsection{Fokker-Planck Equation}

Additional physical insight can be obtained by reorganizing equation~(\ref{eq12})
in the Fokker-Planck form
\begin{equation}
{\partial \Ngreen \over \partial y}
= {\partial^2 \over \partial x^2} \left(
x^q \, \Ngreen\right)
- {\partial \over \partial x} \left[(q+2+a) \, x^{q-1}
\, \Ngreen\right] - \theta \, x^{2-q} \, \Ngreen
+ {\delta(x-x_0) \, \delta(y-y_0)}
\ ,
\label{eq15}
\end{equation}
where we have defined the Green's function number distribution $\Ngreen$
using
\begin{equation}
\Ngreen(x_0,x,y_0,y) \equiv 4 \pi \, m^3 c^3 \, x^2 \, \green(x_0,x,y_0,y)
\ .
\label{eq16}
\end{equation}
The Fokker-Planck coefficients appearing in equation~(\ref{eq15}), which
describe the evolution of the particle distribution due to the influence
of stochastic and systematic processes, are given by \citep{rei65}
\begin{equation}
{1 \over 2}{d\sigma^2 \over dy} = x^q \ , \ \ \ \ \ 
\Big\langle{dx\over dy}\Big\rangle = (q+2+a) \, x^{q-1} \ ,
\label{eq17}
\end{equation}
where the first coefficient describes the ``broadening'' of the distribution
due to momentum space diffusion, and the second represents the mean ``drift''
of the particles (i.e., the average acceleration rate).

Equation~(\ref{eq15}) is equivalent to equation~(49) from \citet{pp95}
if we set their parameters $a_{\rm pp} = 2+a$ and $s_{\rm pp} = q-2$,
where $s_{\rm pp}$ denotes the power-law index describing the momentum
dependence of the escape timescale. Our particular choice for $s_{\rm
pp}$ reflects the physics of the resonant wave-particle interactions, as
represented by equations~(\ref{eq7}). The Fokker-Planck form of
equation~(\ref{eq15}) clearly reveals the fundamental nature of the
transport process. In particular, we note that in the limit $a \to 0$,
the drift coefficient $\langle dx/dy\rangle$ reduces to the purely
stochastic result (eq.~[\ref{eq9}]), as expected when systematic
gains/losses are excluded from the problem. The total number and energy
densities are computed in terms of $\Ngreen$ using (cf. eq.~[\ref{eq5}])
\begin{equation}
n_{_{\rm G}}(y) = \int_0^\infty \Ngreen(x_0,x,y_0,y) \, dx
\ , \ \ \ \ \ 
U_{_{\rm G}}(y) = \int_0^\infty \epsilon \, x \, \Ngreen(x_0,x,y_0,y) \, dx
\ ,
\label{eq18}
\end{equation}
where (see eq.~[\ref{eq3}])
\begin{equation}
\epsilon = m c^2 \left[\left(x^2 + 1\right)^{1/2}-1\right]
\ .
\label{eq19}
\end{equation}
Since the physical situation considered here corresponds to the injection
of a single particle per unit volume at ``time'' $y=y_0$, it follows that
$n_{_{\rm G}}(y_0)=1$.

\section{SOLUTION FOR THE GREEN'S FUNCTION}

The result obtained for the Green's function number distribution
$\Ngreen$ has two important special cases depending on whether $q = 2$
or $q < 2$. \citet{pp95} obtained the exact solution to
equation~(\ref{eq15}) for the hard sphere case \citep[][]{ram79} with
$q=2$ and their parameter $s_{\rm pp}=0$, corresponding to a
momentum-independent escape timescale. In this section we derive the
exact solution to the time-dependent problem with $q < 2$ and $s_{\rm
pp}=q-2$, which describes the physics of the wave-particle interactions
(see eqs.~[\ref{eq7}]).

\subsection{Laplace Transformation}

We define the Laplace transformation of $\Ngreen$ using
\begin{equation}
L(x_0,x,s) \equiv \int_0^\infty e^{-s y} \, \Ngreen(x_0,x,y_0,y) \, dy
\ .
\label{eq20}
\end{equation}
By operating on the Fokker-Planck equation~(\ref{eq15}) with
$\int_0^\infty e^{-s y} \, dy$, we obtain
\begin{equation}
{d^2 \over d x^2} \left(x^q \, L \right)
- {d \over d x} \left[(q+2+a) \, x^{q-1}
\, L \right] - \theta \, x^{2-q} \, L - s L
= - e^{-s y_0} \, \delta(x-x_0)
\ ,
\label{eq21}
\end{equation}
or, equivalently,
\begin{equation}
{d^2 L \over d x^2} + \left({q - 2 - a \over x} \right)
{d L \over d x}
+ \left[{(1-q)(2+a) \over x^2} - {\theta \over x^{2 q-2}}
- {s \over x^q} \right] \, L = - {e^{-s y_0} \, \delta(x-x_0) \over x^q}
\ .
\label{eq22}
\end{equation}
This equation can be transformed into standard form by introducing the
new momentum variables
\begin{equation}
z(x) \equiv {2 \sqrt{\theta} \over 2 - q} \ x^{2-q} \ , \ \ \ \ \ 
z_0(x_0) \equiv {2 \sqrt{\theta} \over 2 - q} \ x_0^{2-q} \ .
\label{eq23}
\end{equation}
After some algebra, we find that the equation for $L$ now becomes
\begin{equation}
z^2 \, {d^2 L \over d z^2} + {a+1 \over q-2} \, z \,
{d L \over d z} + \left[{(1-q)(2+a) \over (2-q)^2}
- {z^2 \over 4} - {s z \over c_0 (2-q)^2}\right] \, L
= - {c_0 \, e^{-s y_0} \, \delta(z-z_0) \over 2-q} \, \left({z \over c_0}
\right)^{(3-2q)/(2-q)}
\ ,
\label{eq24}
\end{equation}
where
\begin{equation}
c_0 \equiv {2 \sqrt{\theta} \over 2-q} \ .
\label{eq25}
\end{equation}
The solutions to equation~(\ref{eq24}) obtained for $z \ne z_0$ that satisfy
the high- and low-energy boundary conditions are given by
\begin{equation}
L(z_0,z,s) = e^{-z/2} \, z^{(a+2)/(2-q)} \,
\cases{
A \, U(\alpha,\beta,z) \ , & $z \ge z_0$ \ , \cr
B \, M(\alpha,\beta,z) \ , & $z \le z_0$ \ , \cr
}
\label{eq26}
\end{equation}
where $M$ and $U$ denote the confluent hypergeometric functions
\citep{as70}, and
\begin{equation}
\alpha \equiv {s + (a+3) \sqrt{\theta} \over 2 (2-q) \sqrt{\theta}}
\ , \ \ \ \ \ \ \
\beta \equiv {a+3 \over 2-q} \ .
\label{eq27}
\end{equation}
The constants $A$ and $B$ appearing in equation~(\ref{eq26}) are determined
by ensuring that the function $L$ is continuous at $z=z_0$, and that it also
satisfies the derivative jump condition
\begin{equation}
\lim_{\varepsilon \to 0} {dL \over dz} \bigg|_{z_0-\varepsilon}
^{z_0+\varepsilon} = - {c_0 \, e^{-s y_0} \over (2-q) \, z_0^2} \,
\left({z_0 \over c_0} \right)^{(3-2q)/(2-q)}
= {- e^{-s y_0} \over 2 \, x_0 \sqrt{\theta}}
\ ,
\label{eq28}
\end{equation}
obtained by integrating the transport equation~(\ref{eq24}) with respect
to $z$ in a small range around the source momentum $z_0$.

The constant $B$ can be eliminated by combining the continuity and derivative
jump conditions. After some algebra, the solution obtained for $A$ is
\begin{equation}
A = - \, {e^{-s y_0} \, e^{z_0/2} \, z_0^{(a+2)/(q-2)} \, M(\alpha,\beta,z_0)
\over 2 \, x_0 \sqrt{\theta} \, W(z_0)}
\ ,
\label{eq29}
\end{equation}
where $W(z)$ denotes the Wronskian, defined by
\begin{equation}
W(z) \equiv
M(\alpha,\beta,z) {d \over dz} U(\alpha,\beta,z)
- U(\alpha,\beta,z) {d \over dz} M(\alpha,\beta,z)
\ .
\label{eq30}
\end{equation}
Using equation~(13.1.22) from \citet{as70}, we find that $W$ is given
by the exact expression
\begin{equation}
W(z) = - \, {\Gamma(\beta) \, z^{-\beta} \, e^z \over
\Gamma(\alpha)}
\ ,
\label{eq31}
\end{equation}
which can be combined with equation~(\ref{eq29}) and the continuity
relation to obtain
\begin{equation}
A = {\Gamma(\alpha) \, z_0^{\beta+(a+2)/(q-2)} \, e^{-s y_0} \, e^{-z_0/2}
\, M(\alpha,\beta,z_0) \over 2 \, x_0 \sqrt{\theta} \, \Gamma(\beta)}
\ ,
\label{eq32}
\end{equation}
\begin{equation}
B = {\Gamma(\alpha) \, z_0^{\beta+(a+2)/(q-2)} \, e^{-s y_0} \, e^{-z_0/2}
\, U(\alpha,\beta,z_0) \over 2 \, x_0 \sqrt{\theta} \, \Gamma(\beta)}
\ .
\label{eq33}
\end{equation}
The final solution for the Laplace transformation $L$ can therefore
be written as
\begin{equation}
L(z_0,z,s) = {\Gamma(\alpha) \, z_0^\beta \over \Gamma(\beta) \,
2 \, x_0 \sqrt{\theta}} \, \left({z \over z_0} \right)^{(a+2)/(2-q)}
\, e^{-s y_0} \, e^{-(z+z_0)/2} \,
M(\alpha,\beta,\zmin) \, U(\alpha,\beta,\zmax)
\ ,
\label{eq34}
\end{equation}
where
\begin{equation}
\zmin \equiv \min(z,z_0) \ , \ \ \ \ \ \ \
\zmax \equiv \max(z,z_0)
\ .
\label{eq35}
\end{equation}

\subsection{Inverse Transformation}

\begin{figure}[t]
\hspace{35mm}
\includegraphics[width=85mm]{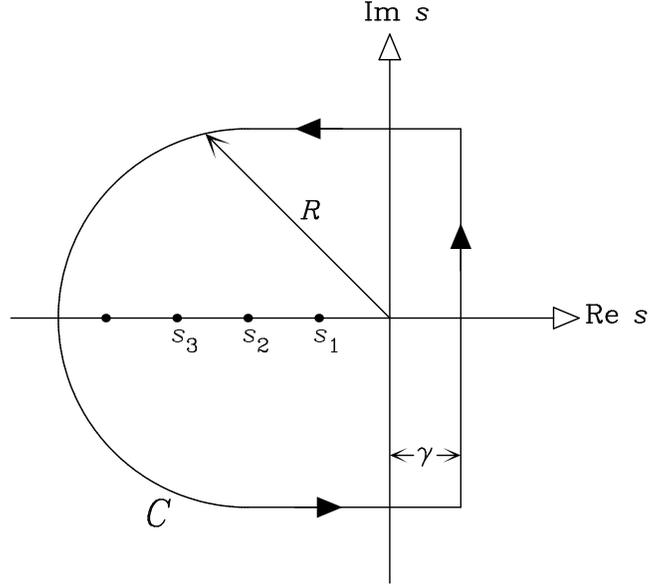}
\caption{Integration contour $C$ used to evaluate equation~(\ref{eq36}).}
\label{fig1}
\end{figure}

The solution for the Green's function $\Ngreen$ can be found using the
complex Mellin inversion integral, which states that
\begin{equation}
\Ngreen(z_0,z,y_0,y) = {1 \over 2 \pi i} \int_{\gamma-i \infty}
^{\gamma + i \infty} e^{s y} \, L(z_0,z,s) \, ds
\ ,
\label{eq36}
\end{equation}
where $\gamma$ is chosen so that the line ${\rm Re} \, s = \gamma$ lies
to the right of any singularities in the integrand. The singularities
are simple poles located where $|\Gamma(\alpha)| \to \infty$, which
corresponds to $\alpha = -n$, with $n = 0,1,2,\ldots$
Equation~(\ref{eq27}) therefore implies that there are an infinite
number of poles situated along the real axis at $s = s_n$, where
\begin{equation}
s_n = \left[2 \, (q-2) \, n - a - 3\right] \sqrt{\theta}
\ , \ \ \ n = 0,1,2,\ldots
\label{eq37}
\end{equation}
We can therefore employ the residue theorem to write
\begin{equation}
\oint e^{s y} \, L(z_0,z,s) \, ds
= 2 \pi i \sum_{n=0}^\infty \ {\rm Res}(s_n)
\ ,
\label{eq38}
\end{equation}
where $C$ is the closed integration contour indicated in Figure~1 and
${\rm Res}(s_n)$ denotes the residue associated with the pole at $s=s_n$.
Based on asymptotic analysis, we conclude that the contribution to the integral
due to the curved portion of the contour vanishes in the limit $R \to \infty$,
and consequently we can combine equations~(\ref{eq36}) and (\ref{eq38}) to obtain
\begin{equation}
\Ngreen(z_0,z,y_0,y) = \sum_{n=0}^\infty \ {\rm Res}(s_n)
\ .
\label{eq39}
\end{equation}
Hence we need only evaluate the residues in order to determine the solution
for the Green's function.

\subsection{Evaluation of the Residues}

The residues associated with the simple poles at $s=s_n$ are evaluated
using the formula \citep[e.g.,][]{but68}
\begin{equation}
{\rm Res}(s_n) = \lim_{s \to s_n}
(s-s_n) \ e^{s y} \, L(z_0,z,s)
\ .
\label{eq40}
\end{equation}
Since the poles are associated with the function $\Gamma(\alpha)$
in equation~(\ref{eq34}), we will need to make use of the identity
\begin{equation}
\lim_{s \to s_n} (s-s_n) \ \Gamma(\alpha)
= {(-1)^n \over n!} \, 2 \, (2-q) \, \sqrt{\theta}
\ ,
\label{eq41}
\end{equation}
which follows from equations~(\ref{eq27}) and (\ref{eq37}). Combining
equations~(\ref{eq34}), (\ref{eq40}), and (\ref{eq41}), we find that the
residues are given by
\begin{equation}
{\rm Res}(s_n) = {(-1)^n \, (2-q) \, e^{s_n (y-y_0)} \, z_0^\beta \over
n! \, \Gamma(\beta) \, x_0} \, \left({z \over z_0}\right)^{(a+2)/(2-q)}
\, e^{-(z+z_0)/2} \, M(-n,\beta,\zmin) \, U(-n,\beta,\zmax)
\ .
\label{eq42}
\end{equation}
Based on equations~(13.6.9) and (13.6.27) from Abramowitz \& Stegun
(1970), we note that the confluent hypergeometric functions appearing in
this expression reduce to Laguerre polynomials, and therefore our result
for the residue can be rewritten after some simplification as
\begin{equation}
{\rm Res}(s_n) = {n! \, e^{s_n (y-y_0)} \, z_0^\beta \, (2-q)
\over \Gamma(\beta+n) \, x_0} \, \left({z \over z_0}\right)^{(a+2)/(2-q)}
\, e^{-(z+z_0)/2} \, P_n^{(\beta-1)}(z) \, P_n^{(\beta-1)}(z_0)
\ ,
\label{eq43}
\end{equation}
where $P_n^{(\beta-1)}(z)$ denotes the Laguerre polynomial.

\subsection{Closed-Form Expression for the Green's Function}

The result for the Green's function number distribution $\Ngreen$ is obtained
by summing the residues (see eq.~[\ref{eq39}]), which yields
\begin{equation}
\Ngreen(z_0,z,y_0,y) = \sum_{n=0}^\infty
\ {n! \, e^{s_n (y-y_0)} \, z_0^\beta \, (2-q)
\over \Gamma(\beta+n) \, x_0} \, \left({z \over z_0}\right)^{(a+2)/(2-q)}
\, e^{-(z+z_0)/2} \, P_n^{(\beta-1)}(z) \, P_n^{(\beta-1)}(z_0)
\ .
\label{eq44}
\end{equation}
This convergent sum is a useful expression for the Green's function.
However, further progress can be made by employing the bilinear
generating function for the Laguerre polynomials, given by
equation~(8.976) from \citet{gr80}. After some algebra, we find that the
general closed-form solution can be written in the form
\begin{equation}
\Ngreen(x_0,x,y_0,y) = {2-q \over x_0} \left({x \over x_0}\right)^{(a+1)/2}
\, {\sqrt{z z_0 \xi} \over 1 - \xi} \ \exp\left[-{(z + z_0)(1+\xi) \over 2
\, (1-\xi)} \right] \, I_{\beta-1}\left({2 \sqrt{z z_0 \xi} \over 1 - \xi}\right)
\ ,
\label{eq45}
\end{equation}
where
\begin{equation}
z(x) \equiv {2 \sqrt{\theta} \over 2 - q} \ x^{2-q} \ , \ \ \ \ \ 
z_0(x_0) \equiv {2 \sqrt{\theta} \over 2 - q} \ x_0^{2-q} \ , \ \ \ \ \ 
\beta \equiv {a+3 \over 2 - q} \ , \ \ \ \ \ 
\xi(y,y_0) \equiv e^{2(q-2)(y-y_0)\sqrt{\theta}}
\ .
\label{eq46}
\end{equation}
Note that the solution for $\Ngreen$ depends on the time parameters $y$
and $y_0$ only through the ``age'' of the injected particles, $y-y_0$, as
indicated by the form for $\xi$. In the limit $\theta \to 0$,
corresponding to infinite escape time, the Green's function number
distribution reduces to
\begin{equation}
\Ngreen(x_0,x,y_0,y) \Big|_{\theta=0} \! = {(x x_0)^{(3-q)/2} \over (2-q)
\, x_0^2 \, (y-y_0)}
\left(x \over x_0 \right)^{a/2} \!\! \exp\left[-{(x^{2-q} + x_0^{2-q}) \over
(2-q)^2 (y-y_0)}\right]
I_{\beta-1}\!\left[{2 (x x_0)^{(2-q)/2} \over (2-q)^2 (y-y_0)}\right]
\ .
\label{eq47}
\end{equation}

The exact solution for the time-dependent Green's function describing the
evolution of a monoenergetic initial spectrum with $q < 2$ given
by equation~(\ref{eq45}) represents an interesting new contribution to
particle transport theory. The corresponding solution for the hard-sphere
case with $q = 2$, given by equation~(43) of \citet{pp95}, can be stated
in our notation as
\begin{eqnarray}
\Ngreen(x_0,x,y_0,y) \Big|_{q=2} = {e^{-\lambda (y-y_0)} \over 2 x_0
\sqrt{\pi (y-y_0)}} \left({x \over x_0}\right)^{(a+1)/2} \, \exp
\left[{-(\ln x - \ln x_0)^2 \over 4 \, (y-y_0)}\right]
\ ,
\label{eq48}
\end{eqnarray}
where
\begin{equation}
\lambda \equiv {(a+1)^2 \over 4} + 2 + a + \theta
\ .
\label{eq49}
\end{equation}
We note that the general solution for $\Ngreen$ given by
equation~(\ref{eq45}) agrees with equation~(\ref{eq48}) in the
limit $q \to 2$ as required. Furthermore, the equation~(\ref{eq45})
allows $q$ to take on {\it negative} values if desired, and it is also
applicable over a broad range of both positive and negative values for
$a$. Recall that when $a=0$, there are no systematic acceleration or
loss processes included in the model. Positive (negative) values for $a$
imply additional systematic acceleration (losses).

\subsection{Transition to the Stationary Solution}

The analytical solutions we have obtained for the Green's function
provide a complete description of the response of the system to the
impulsive injection of monoenergetic particles at any desired time. The
generality of these expressions allows one to obtain the particular
solution for the distribution function $f$ associated with any
arbitrary momentum-time source function $S$ using the convolution
given by equation~(\ref{eq6}). One case of special interest is the
spectrum resulting from the {\it continual injection} of monoenergetic
particles beginning at time $t=0$, described by the source term
\begin{equation}
S(p,t) = \cases{
\dot N_0 \, \delta(p-p_0) \ , & $t \ge 0$ \ , \cr
0 \ , & $t < 0$ \ , \cr
}
\label{eq50}
\end{equation}
where $\dot N_0$ denotes the rate of injection of particles with
momentum $p_0$ per unit volume per unit time. We assume that no
particles are present for $ t < 0$. Combining equations~(\ref{eq6})
and (\ref{eq50}), we find that the time-dependent distribution function
resulting from monoenergetic particle injection is given by
\begin{equation}
f(p,t) = \dot N_0 \int_0^t \green(p_0,p,t_0,t) \, dt_0 \ .
\label{eq51}
\end{equation}
By transforming to the dimensionless variables $x$ and $y$ and employing
equation~(\ref{eq16}), we conclude that the particular solution for the
number distribution associated with continual monoenergetic particle injection
can be written as
\begin{equation}
N(x,y) \equiv 4 \pi \, m^3 c^3 \, x^2 \, f(x,y)
= {\dot N_0 \over D_*} \int_0^y \Ngreen(x_0,x,y_0,y) \, dy_0 \ ,
\label{eq52}
\end{equation}
where we have used equation~(\ref{eq13}) to make the substitution $dt_0
= dy_0 / D_*$. Since $\Ngreen$ depends on the time parameters $y$ and
$y_0$ only through the combination $y-y_0$ (see eqs.~[\ref{eq45}] and
[\ref{eq48}]), it follows that
\begin{equation}
N(x,y) = {\dot N_0 \over D_*} \int_0^y \Ngreen(x_0,x,0,y') \, dy' \ .
\label{eq53}
\end{equation}
This is a more convenient form for $N(x,y)$ because $y$ now appears only
in the upper integration bound.

For general, finite values of $y$, the time-dependent particular
solution for $N(x,y)$ given by equation~(\ref{eq53}) must be computed
numerically by substituting for $\Ngreen$ using either
equation~(\ref{eq45}) or (\ref{eq48}), depending on the value of $q$.
However, as $y \to \infty$, the solution rapidly approaches a stationary
result representing a balance between injection, acceleration, and
particle escape. The form of the stationary solution, called the {\it
steady-state Green's function}, $N^{\rm G}_{\rm ss}$, can be obtained by
directly solving the transport equation~(\ref{eq1}) with $\partial
f/\partial t=0$ and $S(p,t) = \dot N_0 \, \delta(p-p_0)$. Alternatively,
the steady-state Green's function can also be computed by taking the
limit of the time-dependent solution, which yields
\begin{equation}
N^{\rm G}_{\rm ss}(x) \equiv \lim_{y \to \infty}
N(x,y) = {\dot N_0 \over D_*} \int_0^\infty \Ngreen(x_0,x,0,y')
\, dy' \ .
\label{eq54}
\end{equation}
In the $q < 2$ case, we can substitute for $\Ngreen$ using equation~(\ref{eq45})
and evaluate the resulting integral using equation~(6.669.4) from \citet{gr80}.
After some algebra, we obtain
\begin{equation}
N^{\rm G}_{\rm ss}(x) = {\dot N_0 \over (2-q) \, x_0 D_*}
\left(x \over x_0\right)^{(a+1)/2} (x x_0)^{(2-q)/2}
\, I_{\beta-1 \over 2} \left(\sqrt{\theta} \, \xmin^{2-q} \over 2-q\right)
\, K_{\beta-1 \over 2} \left(\sqrt{\theta} \, \xmax^{2-q} \over 2-q\right)
\ ,
\label{eq55}
\end{equation}
where $\beta = (a+3)/(2-q)$, and
\begin{equation}
\xmin \equiv \min(x,x_0) \ , \ \ \ \ \ \ \
\xmax \equiv \max(x,x_0)
\ .
\label{eq56}
\end{equation}
Likewise, for the case with $q=2$, we can substitute for $\Ngreen$ using
equation~(\ref{eq48}) and then utilize equation~(3.471.9) from \citet{gr80}
to conclude that the steady-state solution is given by
\begin{equation}
N^{\rm G}_{\rm ss}(x) \Big|_{q=2} = {\dot N_0 \over 2 \, x_0 D_* \sqrt{\lambda}}
\left(x \over x_0\right)^{(a+1)/2} \left(\xmax \over \xmin\right)^{-\sqrt{\lambda}}
\ ,
\label{eq57}
\end{equation}
where $\lambda$ is defined by equation~(\ref{eq49}). The steady-state
solutions given by equations~(\ref{eq55}) and (\ref{eq57}) agree with
the results obtained by directly solving the transport equation. Due to
the asymptotic behavior of the Bessel $K_\nu(z)$ function for large $z$,
equation~(\ref{eq55}) indicates that $N^{\rm G}_{\rm ss}$ exhibits an
exponential cutoff at high energies when $q < 2$ \citep{as70}. This
corresponds physically to the fact that the escape timescale decreases
with increasing particle momentum in this case. However, when $q=2$
(eq.~[\ref{eq57}]), the spectrum displays a pure power-law behavior at
high energies because the escape timescale is independent of the
particle momentum. Specific examples illustrating these behaviors will be
presented in \S~4.

\subsection{Escaping Particle Distribution}

The various expressions we have obtained for the Green's function
$\Ngreen$ describe the momentum distribution of the particles remaining
in the plasma at time $t$ after injection occurring at time $t_0$.
However, since our model incorporates a physically realistic,
momentum-dependent escape timescale $t_{\rm esc}(p)$ given by
equation~(\ref{eq7}), it is also quite interesting to compute the
spectrum of the {\it escaping} particles, which may form an energetic
outflow capable of producing observable radiation. In general, the
number distribution of the escaping particles, $\dot N^{\rm esc}(x,y)$,
is related to the current distribution of particles in the plasma, $N(x,y)$,
via
\begin{equation}
\dot N^{\rm esc}(x,y) \equiv t_{\rm esc}^{-1} \, N(x,y)
= \theta D_* \ x^{2-q} \, N(x,y)
\ ,
\label{eq58}
\end{equation}
where we have used equations~(\ref{eq7}) and (\ref{eq14}) to obtain the
final result. The quantity $\dot N^{\rm esc}(x,y)\,dx$ represents the number
of particles escaping per unit volume per unit time with dimensionless
momenta between $x$ and $x+dx$.

An important special case is the evolution of the escaping particle
spectrum resulting from {\it impulsive monoenergetic injection} at
dimensionless time $y=y_0$. In this application, equation~(\ref{eq58})
gives the Green's function number spectrum for the escaping particles,
defined by
\begin{equation}
\dot\Ngreen^{\rm esc}(x_0,x,y_0,y)
\equiv \theta D_* \ x^{2-q} \, \Ngreen(x_0,x,y_0,y)
\ ,
\label{eq59}
\end{equation}
where $\Ngreen$ is evaluated using either equation~(\ref{eq45}) or
(\ref{eq48}) depending on the value of $q$. We can use this expression
for $\dot\Ngreen^{\rm esc}$ to compute the {\it total} escaping number
distribution resulting from an impulsive flare occurring at $t=0$ by
integrating the escaping distribution with respect to time, which yields
\begin{equation}
\Delta\Ngreen^{\rm esc}(x) \equiv \int_0^\infty \dot\Ngreen^{\rm esc} \, dt
= \theta \, x^{2-q} \int_0^\infty \Ngreen(x_0,x,0,y') \, dy'
\ ,
\label{eq60}
\end{equation}
where we have used equation~(\ref{eq59}) and taken advantage of the fact
that $\Ngreen$ depends on $y$ and $y_0$ only through the difference $y-y_0$
(cf. eq.~[\ref{eq53}]).
%
%
By comparing equations~(\ref{eq54}) and
(\ref{eq60}), we deduce that
\begin{equation}
\Delta\Ngreen^{\rm esc}(x) = {\theta D_* \over \dot N_0}
\ x^{2-q} \, N^{\rm G}_{\rm ss}(x)
\ ,
\label{eq61}
\end{equation}
so that the total escaping spectrum is simply proportional to the steady-state
Green's function resulting from continual injection, as expected. The process
considered here corresponds to the injection of a single particle at time $t=0$,
and therefore we find that the normalization of $\Delta\Ngreen^{\rm esc}$ is
given by
\begin{equation}
\int_0^\infty \Delta\Ngreen^{\rm esc}(x) \, dx = 1
\ ,
\label{eq62}
\end{equation}
which provides a useful check on the numerical results.

Another interesting example is the case of continual monoenergetic
particle injection commencing at time $t=0$, described by the source
term given by equation~(\ref{eq50}). The time-dependent buildup of the
escaping particle spectrum $\dot N^{\rm esc}(x,y)$ in this scenario can
be analyzed by using equation~(\ref{eq53}) to substitute for $N(x,y)$ in
equation~(\ref{eq58}), which yields
\begin{equation}
\dot N^{\rm esc}(x,y)
= \dot N_0 \, \theta \ x^{2-q} \int_0^y \Ngreen(x_0,x,0,y') \, dy'
\ .
\label{eq63}
\end{equation}
In the limit $y \to \infty$, the escaping particle spectrum approaches the
steady-state result
\begin{equation}
\dot N_{\rm ss}^{\rm esc}(x) \equiv \lim_{y \to \infty} \dot N^{\rm esc}(x,y)
= \dot N_0 \, \theta \ x^{2-q} \int_0^\infty \Ngreen(x_0,x,0,y') \, dy'
\ .
\label{eq64}
\end{equation}
By comparing equations~(\ref{eq54}), (\ref{eq61}), and (\ref{eq64}), we
conclude that
\begin{equation}
\dot N_{\rm ss}^{\rm esc}(x)
= \theta D_* \ x^{2-q} \, N^{\rm G}_{\rm ss}(x)
= \dot N_0 \ \Delta\Ngreen^{\rm esc}(x)
\ .
\label{eq65}
\end{equation}
The final result can be combined with equation~(\ref{eq62}) to show that
the escaping spectrum satisfies the normalization relation
\begin{equation}
\int_0^\infty \dot N_{\rm ss}^{\rm esc}(x) \, dx = \dot N_0 \ ,
\label{eq66}
\end{equation}
as expected for the case of continual steady-state injection.

Note that analytical expressions for the steady-state escaping spectrum $\dot
N_{\rm ss}^{\rm esc}(x)$ can be obtained by substituting for $N^{\rm G}_{\rm
ss}(x)$ in equation~(\ref{eq65}) using either equation~(\ref{eq55}) or
(\ref{eq57}), depending on the value of $q$. We obtain
\begin{equation}
\dot N^{\rm esc}_{\rm ss}(x) = {\dot N_0 \, \theta \, x^{2-q} \over (2-q) \, x_0}
\left(x \over x_0\right)^{(a+1)/2} (x x_0)^{(2-q)/2}
\, I_{\beta-1 \over 2} \left(\sqrt{\theta} \, \xmin^{2-q} \over 2-q\right)
\, K_{\beta-1 \over 2} \left(\sqrt{\theta} \, \xmax^{2-q} \over 2-q\right)
\ ,
\label{eq67}
\end{equation}
for $q < 2$, and
\begin{equation}
\dot N^{\rm esc}_{\rm ss}(x) = {\dot N_0 \, \theta \, x^{2-q} \over
2 \, x_0 \sqrt{\lambda}}
\left(x \over x_0\right)^{(a+1)/2} \left(\xmax \over \xmin\right)^{-\sqrt{\lambda}}
\ ,
\label{eq68}
\end{equation}
for $q = 2$.

\section{RESULTS}

The new result we have derived for the secular Green's function
(eq.~[\ref{eq45}]) displays a rich behavior through its complex
dependence on momentum, time, and the dimensionless parameters $q$, $a$,
and $\theta$, which are related to the fundamental physical transport
coefficients $D_*$, $A_*$, and $t_*$ via equations~(\ref{eq14}). Here we
present several example calculations in order to illustrate the utility
of the new solution. Detailed applications to astrophysical situations,
including active galaxies and $\gamma$-ray bursts, will be presented in
subsequent papers.

\begin{figure}[t]
\includegraphics[width=6in]{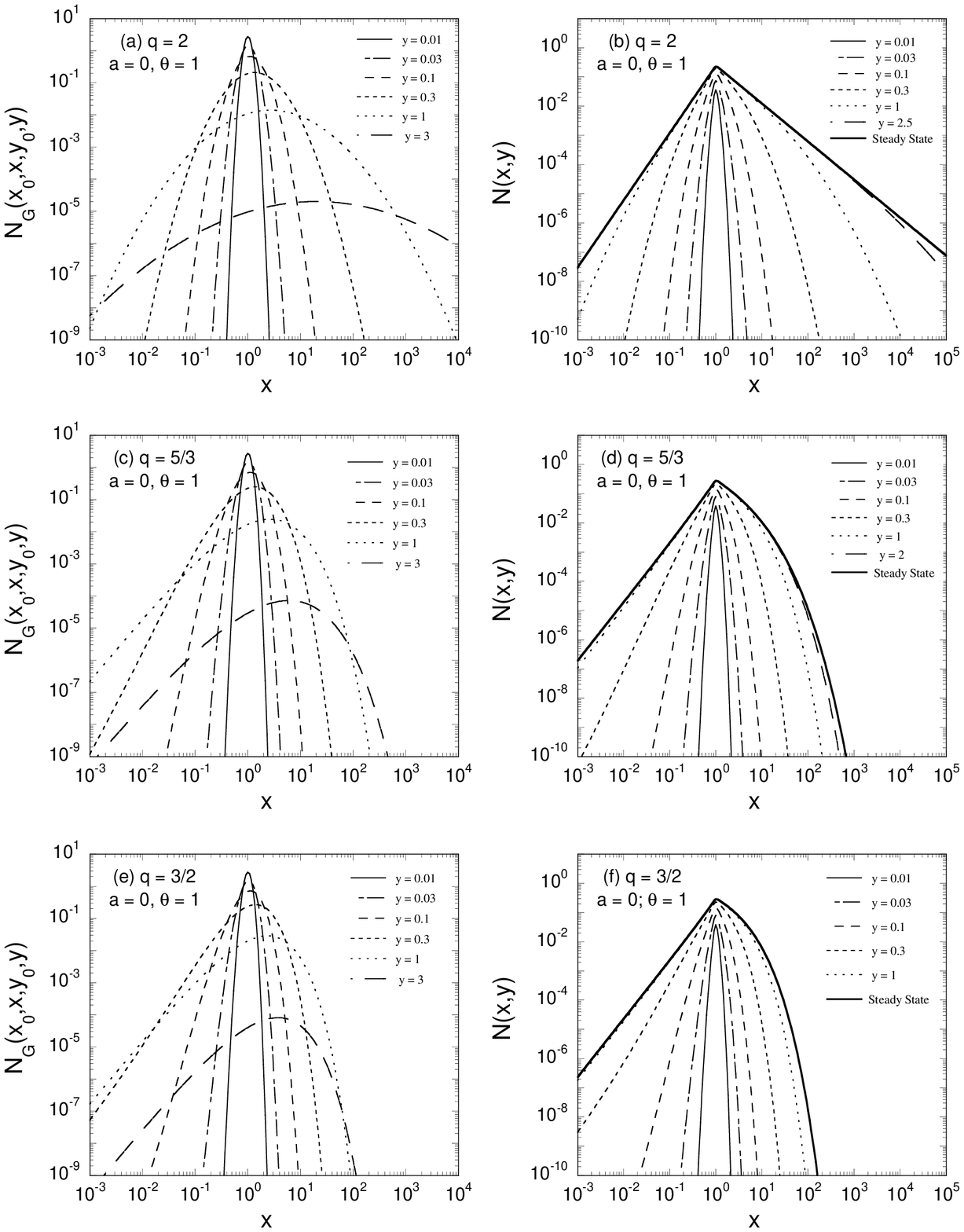}
\vskip-0.6in
\caption{Green's function solutions to the time-dependent stochastic
particle acceleration equation for $x_0=1$. The left panels depict the
impulsive-injection solution, $\Ngreen$ (eqs.~[\ref{eq45}] and
[\ref{eq48}]), and the right panels illustrate the response to uniform,
continuous injection, $N$ (eq.~[\ref{eq53}]), with $\dot
N_0=D_*=1$. Note that $N$ approaches the corresponding steady-state
solution (eqs.~[\ref{eq55}] and [\ref{eq57}]) as $y$ increases. The
indices of the wave turbulence spectra are indicated, with $q = 2$ for
hard-sphere scattering, $q = 5/3$ for a Kolmogorov cascade, and $q =
3/2$ for a Kraichnan cascade.}
\label{fig2}
\end{figure}

The panels on the left-hand side of Figure~2 depict the time-dependent
Green's function solution, $\Ngreen$, describing the evolution of the
particle distribution in the plasma resulting from {\it impulsive}
monoenergetic injection at $y=0$. Results are presented for the
hard-sphere scattering case ($q = 2$), computed using
equation~(\ref{eq48}), and for the Kolmogorov ($q = 5/3$) and Kraichnan
($q = 3/2$) cases, evaluated using equation~(\ref{eq45}). The only
acceleration mechanism considered here is the stochastic acceleration
associated with the second-order Fermi process, and therefore we set $a
= 0$. The escape parameter $\theta$ is set equal to unity, so that the
timescale for escape is comparable to the diffusion timescale. As the
wave turbulence spectrum becomes steeper (i.e., as $q$ increases), a
larger fraction of the turbulence energy is contained in waves with long
wavelengths, which interact resonantly with higher energy particles.
Steeper turbulence spectra therefore produce harder particle
distributions, as can be confirmed in the plots. Consequently we
conclude that an ensemble of hard sphere scattering centers is more
effective at accelerating nonthermal particles compared with a
Kolmogorov wave spectrum, which in turn is more effective than a
Kraichnan spectrum.

The panels on the right-hand side of Figure~2 illustrate the buildup of
the particle spectrum in the plasma, $N(x,y)$, due to {\it continual}
monoenergetic injection beginning at $y=0$, computed by evaluating
numerically the integral in equation~(\ref{eq53}). We have set $a=0$ and
therefore the acceleration is purely stochastic. As $y \to \infty$, the
spectrum approaches the steady-state form given by equation~(\ref{eq55})
for $q<2$ or by equation~(\ref{eq57}) for $q=2$. In the hard-sphere case
($q=2$), the particle spectrum displays a power-law shape at high
energies in agreement with equation~(\ref{eq57}). However, when $q < 2$,
particle escape dominates over acceleration at high energies, and
therefore the steady-state distribution is truncated, even in the
absence of radiative losses. This effect produces the quasi-exponential
turnovers exhibited by the stationary spectra when $q = 5/3$ and $q =
3/2$. Particle escape in these cases mimics energy losses due to, for
example, synchrotron emission from electrons.

\begin{figure}[t]
\hskip0.2in\plottwo{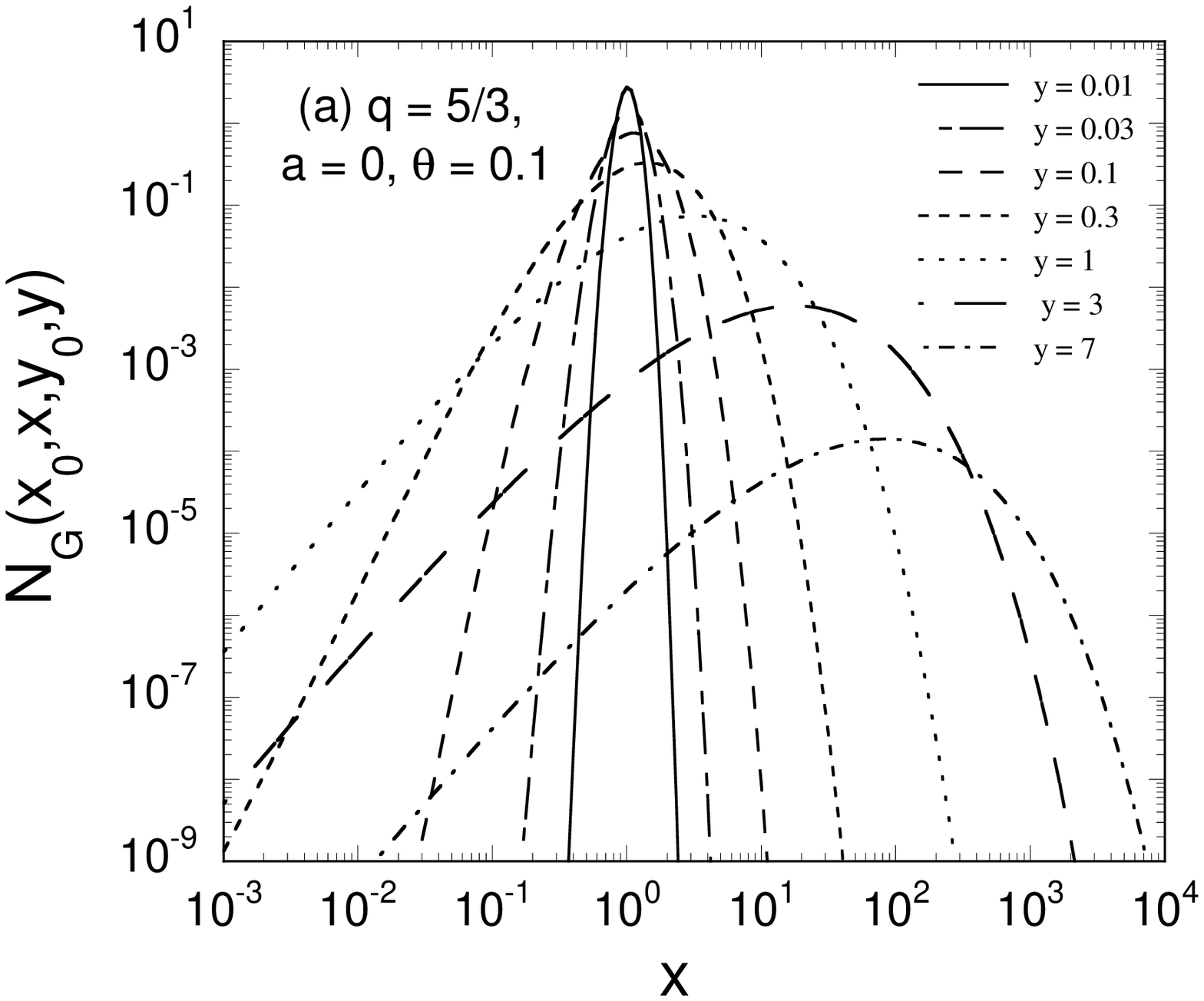}{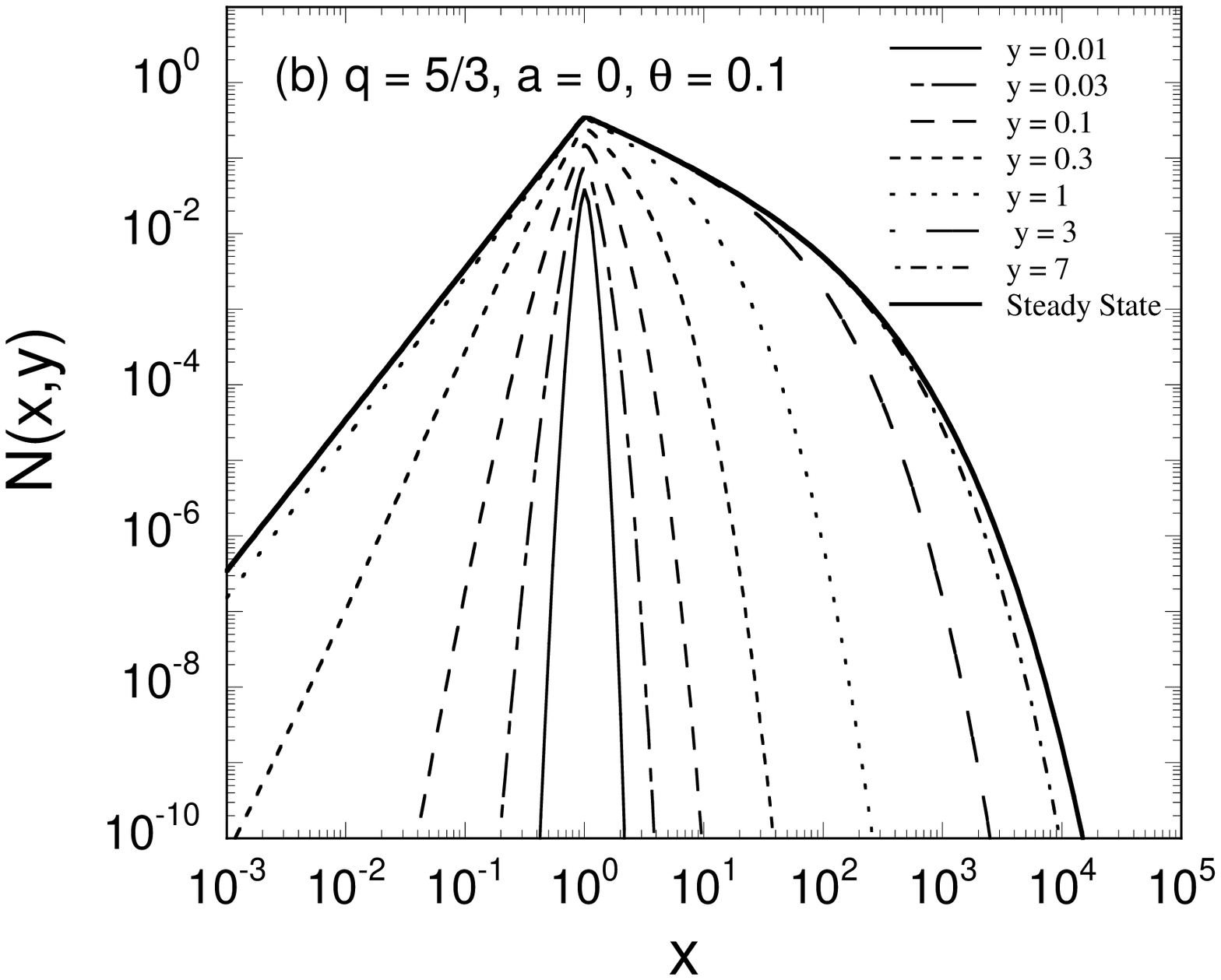}
\hskip0.5truein
\caption{Evolution of the particle distribution in the plasma resulting
from monoenergetic injection with $q = 5/3$, $a = 0$, $\theta = 0.1$,
and $x_0=1$. Panel~({\it a}) depicts the Green's function, $\Ngreen$,
resulting from impulsive monoenergetic injection (eq.~[\ref{eq45}]), and
panel ({\it b}) illustrates the response to continual monoenergetic
injection (eq.~[\ref{eq53}]) and the corresponding steady-state solution
(eq.~[\ref{eq55}]) for $\dot N_0=D_*=1$.}
\label{fig3}
\end{figure}

Figures~3 and 4 illustrate the effects of varying the values of the
escape parameter $\theta$ and the systematic acceleration/loss parameter
$a$ for the $q = 5/3$ case. In Figure~3, we set $a=0$ and $\theta =
0.1$, which represents a particle escape timescale that is an order of
magnitude larger than the corresponding case depicted in Figure~2. The
longer escape time allows the particles to be accelerated to higher mean
energies before diffusing out of the plasma, and the decay of the
particle density at late times takes place much more slowly than for
larger values of $\theta$. The hardening of the spectra causes the
quasi-exponential cutoffs to move to higher energies, and the same
effect can also be noted in the corresponding stationary solutions. The
calculations represented in Figure~4 include additional systematic
acceleration processes that are modeled by setting $a = 2$. The enhanced
particle acceleration further hardens the spectrum and shifts the cutoff
to even higher energies. Although the slope of the low-energy particle
distribution ($x < x_0$) is not altered much when only $\theta$ is
varied (see Figs.~2 and 3), this slope becomes significantly steeper
when $a$ is increased, as indicated in Figure~4.

In the left-hand panels of Figures~5 and 6 we plot the time-dependent
solution for the Green's function describing the escaping particles,
$\dot\Ngreen^{\rm esc}$, resulting from {\it impulsive} monoenergetic
particle injection (eq.~[\ref{eq59}]) when $q = 5/3$. The right-hand
panels illustrate the buildup of the escaping spectrum, $\dot N^{\rm
esc}$ (eq.~[\ref{eq63}]), resulting from {\it continual} injection
beginning at $y=0$. Note the transition to the steady-state solution,
$\dot N_{\rm ss}^{\rm esc}$ (eq.~[\ref{eq64}]), as $y \to \infty$. In
Figure~5 we set $\theta=0.1$ and $a=0$, and in Figure~6 we set
$\theta=0.1$ and $a=2$. Included for comparison are the corresponding
spectra describing the particle distributions in the plasma at the same
values of $y$. We point out that the escaping particle spectra are
significantly harder than the in situ distributions, which reflects the
preferential escape of the high-energy particles resulting from the
momentum dependence of the escape timescale when $q<2$ (see
eq.~[\ref{eq7}]).

\begin{figure}[t]
\hskip0.2in\plottwo{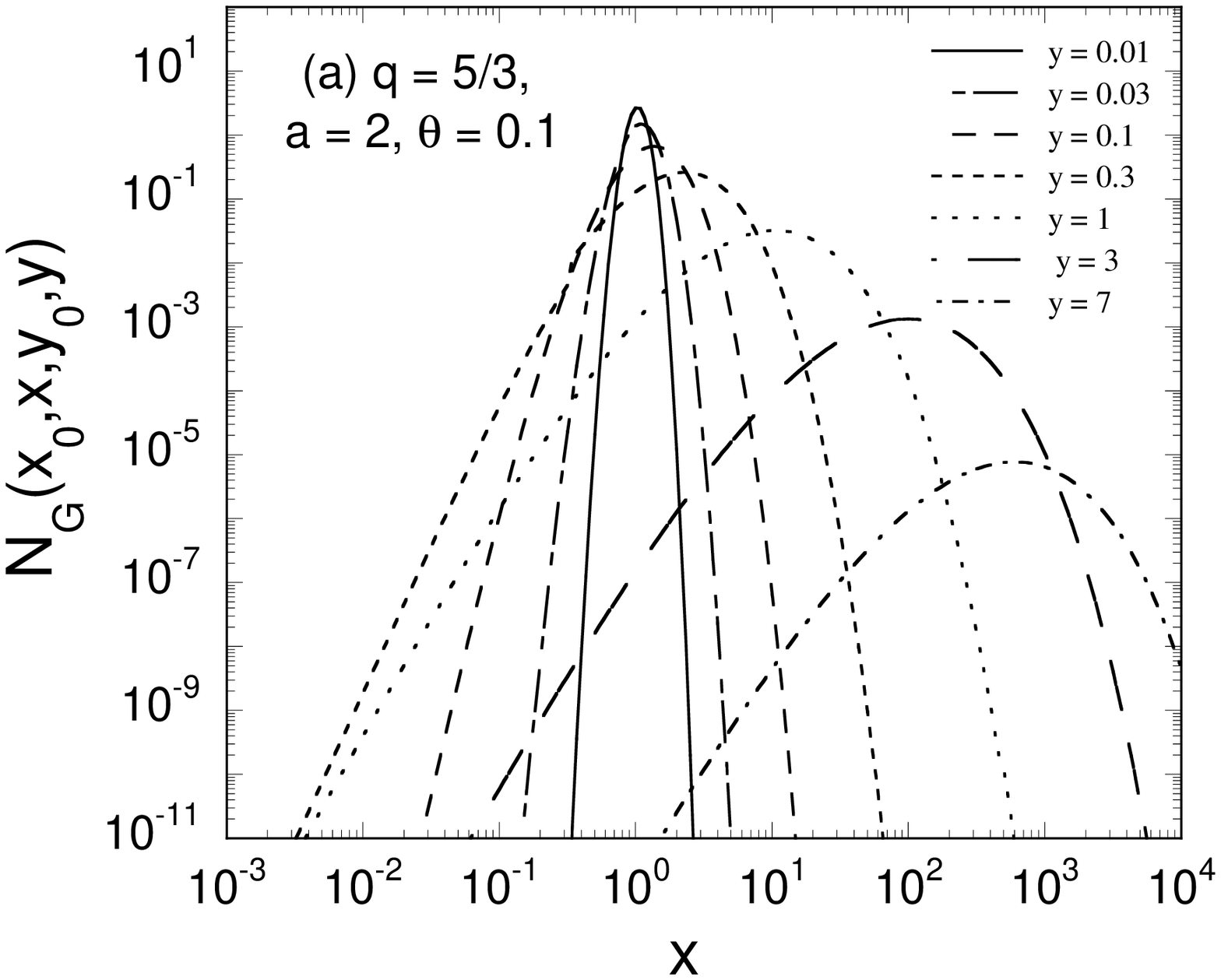}{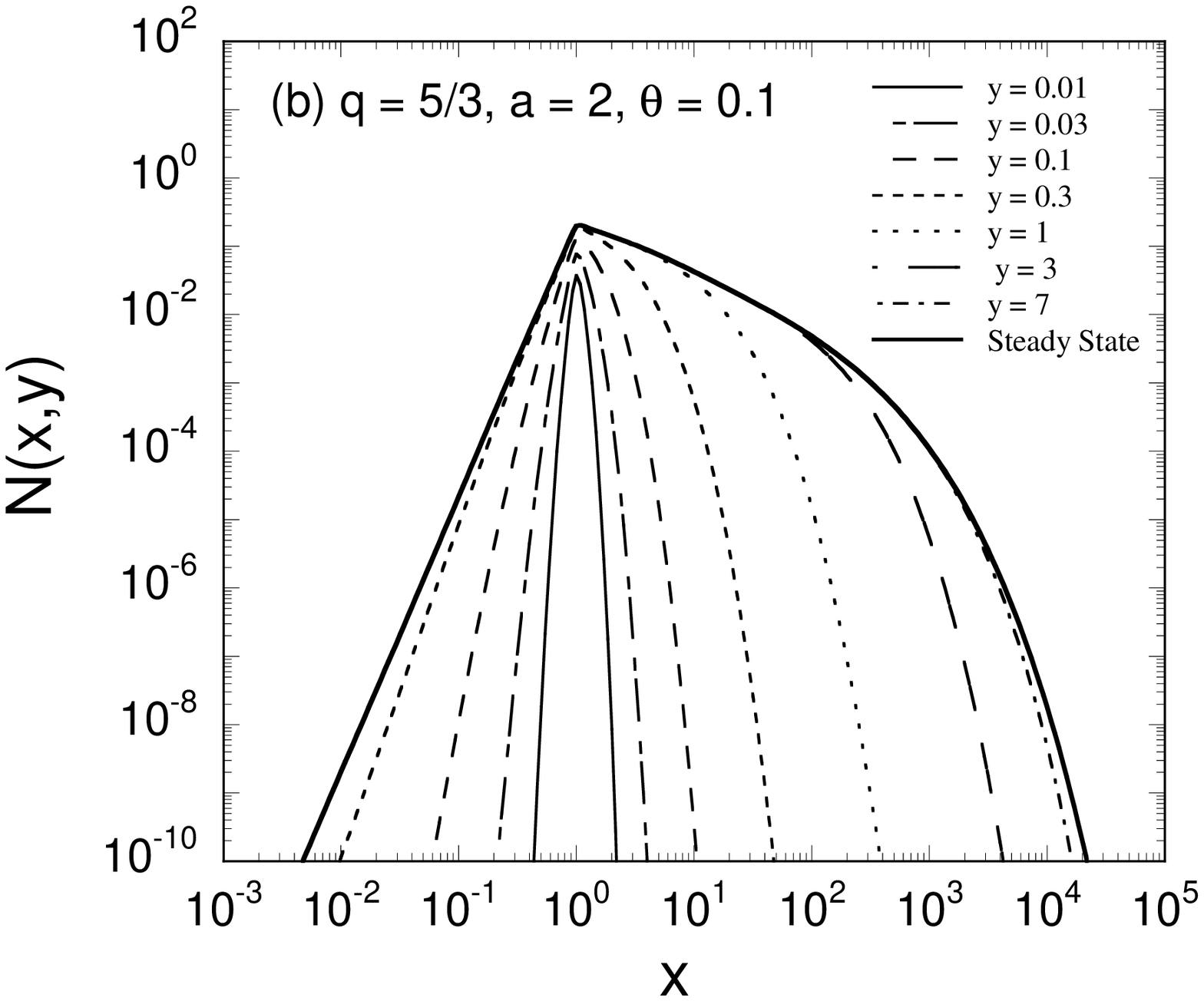}
\caption{Same as Fig.~3, except $a = 2$.}
\label{fig4}
\end{figure}

\section{DISCUSSION AND SUMMARY}

We have derived new, closed-form solutions (eqs.~[\ref{eq45}] and
[{\ref{eq67}]}) for the time-dependent Green's function representing the
stochastic acceleration of relativistic ions interacting with MHD waves.
The analytical results we have obtained describe the time-dependent
distributions for both the accelerated (in situ) and the escaping
particles. The Fokker-Planck transport equation considered here includes
momentum diffusion with coefficient $D(p) \propto p^q$, particle escape
with mean timescale $t_{\rm esc} \propto p^{q-2}$, and additional
systematic acceleration/losses with a rate proportional to $p^{q-1}$,
where $p$ is the particle momentum and $q$ is the index of the wave
turbulence spectrum. This specific scenario describes the resonant
interaction of particles with fast-mode and Alfv\'en waves, which is one
of the fundamental acceleration processes in high-energy astrophysics
\citep[e.g.,][]{dml96}.

The new analytical result complements the work of \citet{pp95} since it
is applicable for any $q < 2$ provided $s_{\rm pp} = q-2$, where $s_{\rm
pp}$ is the power-law index used by these authors to describe the
momentum dependence of the escape timescale. The most closely related
previous result is given by their equation~(59), which treats the case
with $q \ne 2$ and $s_{\rm pp}=0$, corresponding to a
momentum-independent escape timescale. Our analytical solution
(eq.~[\ref{eq45}]) agrees with theirs in the limit $q \to 2$, as
expected. The general features of our new solution were discussed in
\S~4, where it was demonstrated that increasingly hard particle spectra
result from larger values of the wave index $q$, smaller values of the
escape parameter $\theta$, and larger values of the systematic
acceleration parameter $a$. The rich behavior of the solution as a
function of momentum, time, and the parameters $q$, $\theta$, and $a$
provides useful physical insight into the nature of the coupled
energetic/spatial particle transport in astrophysical plasmas.

\begin{figure}[t]
\hskip0.2in\plottwo{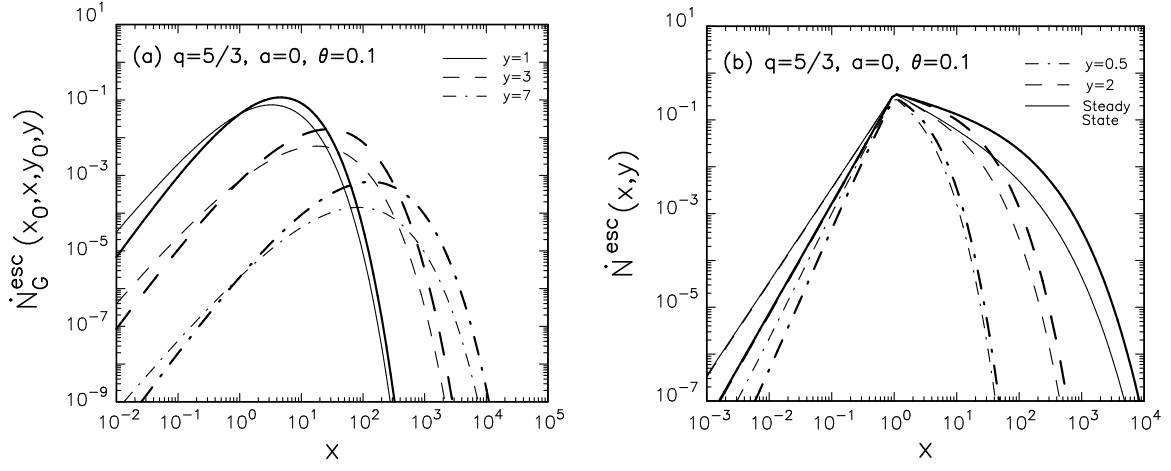}{f5b.eps}
\caption{Evolution of the particle distribution resulting from
monoenergetic injection with $q = 5/3$, $a = 0$, $\theta = 0.1$, $x_0 =
1$, $\dot N_0=1$, and $D_*=1$. Panel~({\it a}) treats the case of
impulsive injection, with the thin lines representing the particle
distribution in the plasma (eq.~[\ref{eq45}]) and the heavy lines denoting
the escaping particle spectrum (eq.~[\ref{eq59}]) in units of $\theta$.
Panel~({\it b}) illustrates the response to continual monoenergetic
injection for the particle distribution in the plasma (eqs.~[\ref{eq53}] and
[\ref{eq55}]; thin lines) and for the escaping particle spectrum
(eqs.~[\ref{eq63}] and [\ref{eq65}]; heavy lines).
See the discussion in the text.}
\label{fig5}
\end{figure}

\begin{figure}[t]
\hskip0.2in\plottwo{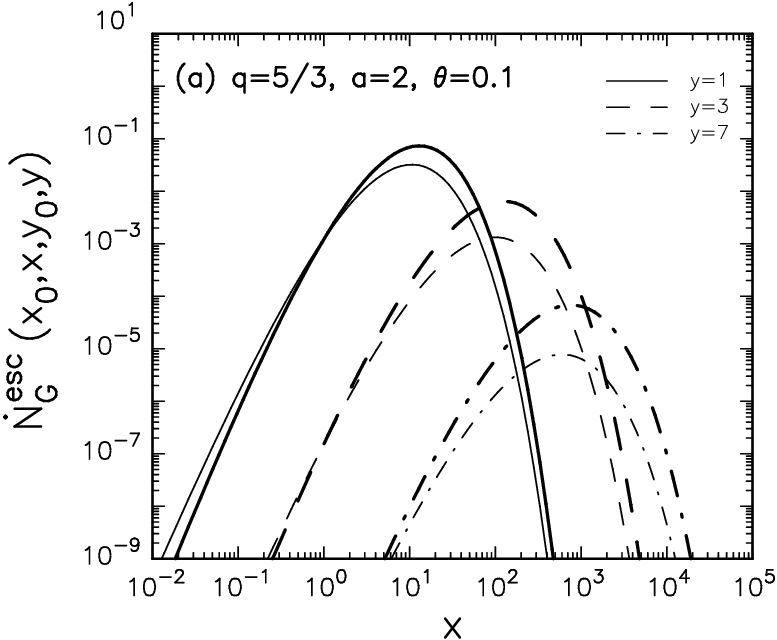}{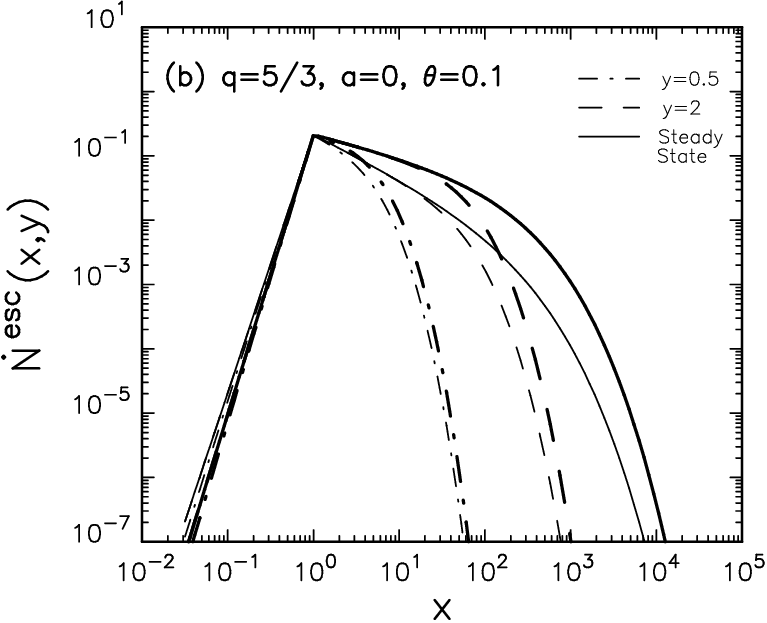}
\caption{Same as Fig.~5, except $a = 2$.}
\label{fig6}
\end{figure}

The solutions presented here can be used to describe the acceleration
and transport of relativistic ions in astrophysical environments in
which the turbulence spectrum is very poorly known and can be
approximated by a power law, such as $\gamma$-ray bursts, active
galaxies, magnetized coronae around black holes, and the intergalactic
medium in clusters of galaxies. For example, the hard X-ray emission
from black-hole jet sources such as Cygnus X-1 \citep{mal06} and the
microquasar LS 5039 \citep{aha05} could be powered by the stochastic
acceleration of ions in a black-hole accretion disk corona that
subsequently escape and interact with surrounding material. In this
scenario, persistent acceleration of monoenergetic particles injected
into the corona, or flaring episodes averaged over a sufficiently long
time, would produce a time-averaged escaping distribution of
relativistic protons given by equation~(\ref{eq67}) for $q < 2$.
Assuming only stochastic acceleration, so that $a = 0$, the number
distribution of escaping particles with $x \geq x_0$ takes the form
\begin{equation}
\dot N_{\rm ss}^{\rm esc}(x) 
\propto x^{(7-3q)/2} K_{{\beta - 1\over 2}}\left( {\sqrt{\theta} 
x^{2-q} \over 2-q}\right) \propto \cases{
x^{3-2q}  \ , & $x \ll \left( {2-q\over \sqrt{\theta}}\right)^{1/(2-q)}$ \  \cr
x^{(5-2q)/2}\exp\left(-{\sqrt{\theta}x^{2-q}\over 2-q}\right)  \ , & 
$x \gg \left({2-q\over \sqrt{\theta}}\right)^{1/(2-q)}$ \  \cr
}.
\label{eq69}
\end{equation}
When the escaping hadrons collide with ambient gas or stellar wind
material, they would generate X-rays and $\gamma$-rays via a pion
production cascade with a very hard spectrum leading up to a
quasi-exponential cutoff. The {\it Gamma ray Large Area Space Telescope}
(GLAST) telescope will be able to provide detailed spectra from Galactic
black-hole sources and unidentified EGRET $\gamma$-ray sources to test
for the existence of this hard component. Our new analytical solution
can also be used to model the variability of radiation from ions
accelerated in the accretion-disk coronae of Seyfert galaxies by
changing the level of turbulence. Flaring $\sim 100\,$MeV -- GeV
radiation could be weakly detected by GLAST as a consequence of this
process. Additional applications of our work include studies of the
stochastic acceleration of relativistic cosmic rays in $\gamma$-ray
bursts \citep{dh01}.

The results we have obtained describe both the momentum distribution of
the particles in the plasma, and the momentum distribution of the
particles that escape to form energetic outflows. Since the solutions do
not include inverse-Compton or synchrotron losses, which are usually
important for energetic electrons, they are primarily applicable to
cases involving ion acceleration. In order to treat the acceleration of
relativistic electrons, a generalized calculation including both
stochastic particle acceleration and radiative losses is desirable, and
we are currently working to extend the analytical model presented here
to incorporate these effects. Beyond the direct utility of the new
analytical solutions for probing the nature of particle acceleration in
astrophysical sources, we note that they are also useful for
benchmarking numerical simulations.

\acknowledgements
T.~L. is funded through NASA {\it GLAST} Science Investigation No.~DPR-S-1563-Y.
The work of C.~D.~D. and P.~A.~B. is supported by the Office of Naval Research.
The authors also acknowledge the useful comments provided by the anonymous
referee.

\appendix

\section{APPENDIX}

In this section we explore the relationship between the momentum diffusion
coefficient $D(p)$ and the mean rate of change of the particle momentum and
kinetic energy, using
the integral method outlined by \citet[][]{sbk99}.

\subsection{Mean Rate of Change of Momentum Due to Stochastic
Acceleration}

In the case of pure stochastic acceleration, the transport
equation~(\ref{eq1}) reduces to
\begin{equation}
{\partial f \over \partial t} = {1 \over p^2} {\partial \over \partial p}
\left[p^2 \, D(p) \, {\partial f \over \partial p}\right] \ .
\label{appen1}
\end{equation}
The mean momentum, $\langle p \rangle$, is defined as a function of $t$ by
\begin{equation}
\langle p \rangle \equiv {1 \over n} \int_0^\infty 4 \pi p^3
\, f(p,t) \, dp \ ,
\label{appen2}
\end{equation}
where the number density $n$ is related to $f$ via equation~(\ref{eq2}).
The associated mean rate of change of the momentum is given by
\begin{equation}
\Big\langle {dp \over dt} \Big\rangle
= {1 \over n} \int_0^\infty 4 \pi p^3 \, {\partial f \over \partial t}
\, dp \ .
\label{appen3}
\end{equation}
Combining this relation with the transport equation~(\ref{appen1}) yields
\begin{equation}
\Big\langle {dp \over dt} \Big\rangle
= {1 \over n} \int_0^\infty 4 \pi p \, {\partial \over \partial p}
\left(p^2 D {\partial f \over \partial p}\right) dp \ .
\label{appen4}
\end{equation}
Integrating by parts once gives
\begin{equation}
\Big\langle {dp \over dt} \Big\rangle
= - {1 \over n} \int_0^\infty 4 \pi p^2 D {\partial f \over \partial p}
\, dp \ ,
\label{appen5}
\end{equation}
and integrating by parts a second time yields
\begin{equation}
\Big\langle {dp \over dt} \Big\rangle
= {1 \over n} \int_0^\infty 4 \pi f {d \over dp}\left(p^2 D \right)
dp \ .
\label{appen6}
\end{equation}
It is sufficient for our purposes to consider the evolution of the
particle distribution $f$ satisfying the monoenergetic initial condition
\begin{equation}
f(p,t) \bigg|_{t=t_0} = A_0 \, \delta(p-p_0) \ .
\label{appen7}
\end{equation}
Combining this relation with equation~(\ref{appen6}), we find that
at the initial time $t=t_0$,
\begin{equation}
\Big\langle {dp \over dt} \Big\rangle \Bigg|_{t=t_0}
= {1 \over p^2} {d \over dp} \left(p^2 D \right) \Bigg|_{p=p_0} \ .
\label{appen8}
\end{equation}
Dropping the subscript ``0'' without loss of generality, we conclude that
the mean rate of change of the momentum for particles
with momentum $p$ at time $t$ is given by
\begin{equation}
\Big\langle {dp \over dt} \Big\rangle
= {1 \over p^2} {d \over dp} \left(p^2 D \right) \ ,
\label{appen9}
\end{equation}
which agrees with equation~(\ref{eq8}).

\subsection{Mean Rate of Change of Kinetic Energy Due to Stochastic
Acceleration}

The mean kinetic energy, $\langle \epsilon \rangle$, is defined by
\begin{equation}
\langle \epsilon \rangle \equiv {1 \over n} \int_0^\infty 4 \pi p^2
\epsilon \, f \, dp \ ,
\label{appen10}
\end{equation}
where $\epsilon$ is given by equation~(\ref{eq3}).
The associated mean rate of change is given by
\begin{equation}
\Big\langle {d\epsilon \over dt} \Big\rangle
= {1 \over n} \int_0^\infty 4 \pi p^2 \epsilon \, {\partial f \over \partial t}
\, dp \ ,
\label{appen11}
\end{equation}
which can be combined with the transport equation~(\ref{appen1}) to obtain
\begin{equation}
\Big\langle {d\epsilon \over dt} \Big\rangle
= {1 \over n} \int_0^\infty 4 \pi \epsilon \, {\partial \over \partial p}
\left(p^2 D {\partial f \over \partial p}\right) dp \ .
\label{appen12}
\end{equation}
As in the previous section, we integrate by parts to find that
\begin{equation}
\Big\langle {d\epsilon \over dt} \Big\rangle
= - {1 \over n} \int_0^\infty 4 \pi {d\epsilon \over dp} \, p^2 D
{\partial f \over \partial p} \, dp \ .
\label{appen13}
\end{equation}
Next we employ the derivative relation
\begin{equation}
{d\epsilon \over dp} = v
\label{appen14}
\end{equation}
and integrating by parts again to obtain
\begin{equation}
\Big\langle {d\epsilon \over dt} \Big\rangle
= {1 \over n} \int_0^\infty 4 \pi f {d \over dp}\left(v p^2 D \right)
dp \ .
\label{appen15}
\end{equation}
Applying the initial condition given by equation~(\ref{appen7}), we find that
at the initial time $t=t_0$
\begin{equation}
\Big\langle {d\epsilon \over dt} \Big\rangle \Bigg|_{t=t_0}
= {1 \over p^2} {d \over dp} \left(v p^2 D \right) \Bigg|_{p=p_0} \ .
\label{appen16}
\end{equation}
Dropping the subscript ``0'' without loss of generality, we conclude that
the mean rate of change of the kinetic energy for particles
with momentum $p$ at time $t$ is given by
\begin{equation}
\Big\langle {d\epsilon \over dt} \Big\rangle
= {1 \over p^2} {d \over dp} \left(p^2 v D \right) \ ,
\label{appen17}
\end{equation}
which agrees with equation~(\ref{eq8b}) and with equation~(13)
from Miller \& Ramaty (1989).


\end{document}